\documentclass[preprint,aps,floatfix,showpacs]{revtex4}
\usepackage{graphicx}
\newcommand{\fr}{\frac}
\newcommand{\beq}{\begin{equation}}
\newcommand{\eeq}{\end{equation}}
\newcommand{\bea}{\begin{eqnarray}}
\newcommand{\eea}{\end{eqnarray}}
\begin{document}
\preprint{\vbox{\hbox{JLAB-THY-06-601} }}
\vspace{0.5cm}
\title{\phantom{x}
\vspace{0.5cm} Photo-production  of Positive Parity Excited Baryons
in the $1/N_c$ Expansion of QCD}

\author{
J. L. Goity $^{a,b}$
\thanks{e-mail: goity@jlab.org} and
N. N. Scoccola $^{c,d,e}$
\thanks{e-mail: scoccola@tandar.cnea.gov.ar}}

\affiliation{
$^a$ Department of Physics, Hampton University, Hampton, VA 23668, USA. \\
$^b$  Thomas Jefferson National Accelerator Facility, Newport News, VA 23606, USA. \\
$^c$ Physics Depart., Comisi\'on Nacional de Energ\'{\i}a
At\'omica,
     (1429) Buenos Aires, Argentina.\\
$^d$ CONICET, Rivadavia 1917, (1033) Buenos Aires, Argentina.\\
$^e$ Universidad Favaloro, Sol{\'\i}s 453, (1078) Buenos Aires,
Argentina.}
\date{\today}
\begin{abstract}
We analyze the photo-production helicity amplitudes for the
positive parity excited baryons in the context of the $1/N_c$
expansion of QCD. The results show that sub-leading corrections in
$1/N_c$ are important  and that, while 1-body effective
operators are dominant,  there is  some evidence for the need
of 2-body effects.
\end{abstract}
\maketitle

This letter addresses the photo-production helicity amplitudes of
the   positive parity excited baryons belonging to the
$[20,\, \ell^+]$ of $SU(4)\times O(3)$ with $\ell=0$ and $2$ in the
context of the $1/N_c$ expansion.  Baryon photo-couplings have been
the subject of many studies over a period of forty years and are
 key elements  in  the understanding of baryon
structure and dynamics  (for a recent review see
Ref.\cite{Lee:2006xu}). The most commonly used tools of analysis
have been the different versions of the constituent quark model
(CQM) \cite{Capstick:2000qj}, and also the so-called  \lq\lq
single-quark transition theory\rq\rq  (SQTT) based on $SU(6)_W$
symmetry \cite{Clo79}.  An approach consistent with QCD is
implemented through the $1/N_c$ expansion \cite{tHo74}.
A variety of  phenomenological applications in the baryon sector, in particular
to the  spectrum and to strong and electromagnetic transitions,
indicate  that  the $1/N_c$ expansion can be a useful tool of
analysis \cite{Man98}.  In   photo-production,
some model independent relations have been obtained in
Ref.\cite{Cohen:2004bk},  while in Ref.\cite{Carlson:1998si}  the case of negative parity  baryons  belonging to the mixed-symmetric spin-flavor multiplet  $[20',\, 1^-]$ were studied at leading order.
In this work we perform the
analysis including ${\cal{O}}(1/N_c)$ corrections to the
mentioned helicity amplitudes of positive parity baryons.

In the large $N_c$ and  isospin  symmetry limits,  a dynamical
contracted spin-flavor $SU(4)$ symmetry  emerges both for ground state
\cite{Gervais} and excited baryons \cite{Pirjol}. A framework for
implementing the latter is with states filling multiplets of $SU(4)\times O(3)$ \cite{Goity}, which is convenient because of the
weakness of the ${\cal{O}}(N_c^0)$ breaking effects  shown by the
excited baryon spectrum.  In this work
that framework is followed, which is particularly simple for the $SU(4)$ symmetric multiplets  involved here.  For details on the definition of states relevant
to this work see e.g. Ref.\cite{Goity:2005rg}.

The helicity amplitudes of interest are defined in the  standard
form \cite{PDG06} which includes a sign factor $\eta(B^*)$ from the strong
amplitude of the decay of the excited baryon to a $\pi N$ state.
They are  given by:
\beq
A_\lambda = - \sqrt{\fr{2\pi\alpha}{\omega}}\;
\eta(B^*)
  \langle B^*, \lambda \mid
{\vec\epsilon}_{+1}\cdot\vec{J}(\omega \hat{z})
\mid N, \lambda-1 \rangle
\label{uno}
\eeq
where $\lambda=1/2$ or $3/2$ is the helicity defined
along  the $\hat{z}$-axis  which  coincides with the photon
momentum,  ${\vec\epsilon}_{+1}$ is the photon's polarization
vector for helicity $+1$,  and $\alpha$ is the
fine-structure constant, and $\mid N\rangle$  denotes
the initial nucleon state.  The electromagnetic current $\vec{J}$ is
represented as a linear combination of effective current operators
which have the most general form:
\bea
\left(k_\gamma^{[L']}  {\cal B}^{[LI]}\right)^{[1I]},
\eea
where the upper scripts  display  the angular momentum and isospin,
and throughout the neutral component, i.e.  $I_3=0$, is taken.
$k_\gamma^{[L']} $ is an irreducible tensor in terms of the photon
momentum, chosen here to be a spherical harmonic,  and  ${\cal
B}^{[L I]}=\left(\xi^{(\ell)} {\cal{G}}^{[\ell' I]}\right)^{[LI]}$
are operators where  $ \xi^{(\ell)}$ is  the $O(3)$  tensor associated with
the excited baryon state normalized by its reduced matrix element  $\langle 0 \mid\mid \xi^{(\ell)}\mid\mid \ell\rangle= \sqrt{2\ell+1}$, and   ${\cal{G}}^{[\ell' I]}$ is a
spin-flavor tensor operator with   $I=0$ or $1$ that can be expressed in terms of products of the  generators of $SU(4)$, namely $S$, $T$ and $G$.  In the case of the Roper
multiplet $[20,\, 0^+]$ there are magnetic dipole and electric
quadrupole transitions for which $L'=1$ in the first case and
$L'=1$  or 3 in the second. For the $[20,\, 2^+]$,  $L'$ can
be 1, 3 or 5.  For general $N_c$ the isovector and isoscalar
components of the electric charge can be generalized in different
ways\cite{Lebed:2004fj}.
Here we consider them as being both of ${\cal O}(N_c^0)$,
corresponding to the  assumption that quark charges
are $N_c$ independent .
Bases of effective operators ${\cal B}^{[L I]}$ can be obtained
along similar steps to those followed, for instance, in the study
of strong transitions \cite{Goity:2005rg}.  Since there is a one
to one correspondence  between $L$ and the multipole to which an
operator contributes to ($M_L$ if $L=1,3$, $E_L$ if $L=2,4$), we
denote them accordingly,  e.g., $E_n^{[LI]}$ is the $n^{th}$  $E_L$ operator of isospin $I$.  The bases of
operators and the order of their matrix elements in $1/N_c$ are
given in Tables~I and~II. Note that in the  case of the $[20, \, 0^+]$ multiplet   we can simply replace  $\xi^{(\ell=0)}=1$.
\begin{table}[h]
\begin{center}
\caption{Basis operators for the
photo-production amplitudes of the non-strange $[20, \, 0^+]$
baryons.}\vspace*{-.0cm}
\begin{tabular}{clc}
\hline \hline
$n$-bodyness &  \hspace*{.5cm}  Operators      & $1/N_c$ order  \\
\hline
 1B
     & $M^{[10]}_1 =\frac{1}{N_c} \frac{1}{\sqrt{3}}\; S^{[1,0]}$
     &  $1/N_c$      \\
     & $M^{[11]}_1 =\frac{1}{N_c} \sqrt{\frac{9}{8}}\; G^{[1,1]}$
     &  $1$      \\
 2B
     & $M^{[11]}_2 =\frac{1}{N_c^2} \frac{1}{2}\;\left[ S , G \right] ^{[1,1]}$
     &  $1/N_c$  \\
     & $E^{[21]}_1 =\frac{1}{N_c^2}\frac{3}{2\sqrt{5}} \;  \left\{ S , G \right\} ^{[2,1]}$
     &  $1/N_c$      \\
\hline \hline\end{tabular}
\end{center}
\end{table}
\vspace*{-.3cm}
The numerical factors in front of each operator conveniently normalize the largest reduced matrix  element  (defined in Eq.(\ref{RME}) below) to be equal to  1 (1/3) for
operators   ${\cal O}(1)$ (${\cal O}(1/N_c))$   when  $N_c=3$.
\vspace*{-.3cm}
\begin{table}[h]
\begin{center}
\caption{Basis operators for the
photo-production amplitudes of the non-strange $[20, \, 2^+]$
baryons. $B$ is the baryon number operator.}\vspace*{-.3cm}
\begin{tabular}{clc}
\hline \hline
$n$-bodyness   &  \hspace*{.5cm}  Operators      & $1/N_c$ order  \\
\hline
 1B
     & $M^{[10]}_1 =  \frac{1}{N_c}\frac{1}{\sqrt{3}}\; \left( \xi^{(2)}\; S\right)^{[1,0]}$
     &  $1/N_c$      \\
     & $M^{[11]}_1 =  \frac{1}{N_c} \frac{6}{5}\;\left( \xi^{(2)}\; G\right)^{[1,1]}$
     &  $1$      \\
2B
     & $M^{[11]}_2 =  \frac{1}{N_c^2} \frac{1}{\sqrt{2}}\; \left( \xi^{(2)} \;\left[ S , G \right] ^{[1,1]}\right)^{[1,1]}$
     &  $1/N_c$  \\
     & $M^{[11]}_3 =  \frac{1}{N_c^2}\sqrt{\frac{3}{2}}\; \left( \xi^{(2)} \; \left\{ S , G \right\}^{[2,1]}\right)^{[1,1]}$
     &  $1/N_c$  \\
1B
     & $E^{[20]}_1 =  \frac{1}{N_c}\frac{1}{2\sqrt{3}}\; \left( \xi^{(2)}\; B\right)^{[2,0]}$
     &  $1$      \\
     & $E^{[20]}_2 =  \frac{1}{N_c}\frac{1}{\sqrt{3}}\; \left( \xi^{(2)}\; S\right)^{[2,0]}$
     &  $1/N_c$      \\
     & $E^{[21]}_1 =  \frac{1}{N_c}\frac{1}{3} \;\left( \xi^{(2)}\; T\right)^{[2,1]}$
     &  $1/N_c$      \\
     & $E^{[21]}_2 =  \frac{1}{N_c} \sqrt{\frac{27}{28}}\;\left( \xi^{(2)} \;G\right)^{[2,1]}$
     &  $1$      \\
2B
     & $E^{[21]}_3 =  \frac{1}{N_c^2}  \sqrt{\frac{3}{14}}\;\left( \xi^{(2)}\;\left[ S , G \right] ^{[1,1]}\right)^{[2,1]}$
     &  $1/N_c$  \\
     & $E^{[21]}_4 =  \frac{1}{N_c^2}\frac{3}{\sqrt{14}}\;  \left( \xi^{(2)}\;\left\{ S , G \right\}^{[2,1]}\right)^{[2,1]}$
     &  $1/N_c$  \\
1B
     & $M^{[30]}_1 =  \frac{1}{N_c}\frac{1}{\sqrt{7}} \;\left( \xi^{(2)}\; S\right)^{[3,0]}$
     &  $1/N_c$      \\
     & $M^{[31]}_1 =  \frac{1}{N_c} \frac{3}{4}\;\left( \xi^{(2)}\; G\right)^{[3,1]}$
     &  $1$      \\
2B
     & $M^{[31]}_2 =  \frac{1}{N_c^2} \frac{1}{\sqrt{8}}\; \left( \xi^{(2)} \;\left[ S , G \right] ^{[1,1]}\right)^{[3,1]}$
     &  $1/N_c$  \\
     & $M^{[31]}_3 =  \frac{1}{N_c^2} \sqrt{\frac{3}{8}}\; \left( \xi^{(2)} \;\left\{ S , G \right\} ^{[2,1]}\right)^{[3,1]}$
     &  $1/N_c$  \\
     & $E^{[41]}_1 =  \frac{1}{N_c^2} \frac{1}{2}\; \left( \xi^{(2)}\; \left\{ S , G \right\}^{[2,1]}\right)^{[4,1]}$
     &  $1/N_c$  \\
\hline \hline
\end{tabular}
\end{center}
\end{table}

\vspace*{-.5cm}
We have checked that the 1-body operators can
be put in direct correspondence with the operators that appear in
old  $SU(6)_W$ analyses \cite{GK74,Babcock:1975bw}. 
The E- and M-multipole
components of a given helicity amplitude of isospin I can be expressed in terms
of the reduced matrix elements (RME) of the operators ${\cal B}^{[LI]}$
as follows:
\bea
&& \!\!\!\!\!\!\!\!\!\!\!\!\!\!\!\!
A_\lambda^{X^{[LI]}}(I_3, J^* I^*) =
\fr{(-1)^{J^*+I^*+I+1}\ w_X(L)\;\eta(B^*)\;}{\sqrt{(2 J^* + 1)(2 I^*+1) }}\;
 \nonumber\\
&\times &  \sqrt{\fr{3 \alpha N_c}{4\omega}} \
\langle L , 1; \fr 12 , \lambda-1 \mid J^* , \lambda \rangle
\langle I , 0; \fr 12 , I_3 \mid I^* , I_3 \rangle
\nonumber \\
&\times&
\sum_{a} g_a(\omega) \
\langle  J^* I^*\parallel {\cal B}_a^{[LI]} \parallel \fr 1 2 \rangle
\eea
where
$X=M\ (E)$ for $L={\rm odd}\ ({\rm even})$,
and $w_X(L)=1\ (\sqrt{(L+1)/(2L+1)})$ if $X=M\  (E)$. $I_3$
denotes the initial nucleon's isospin.
The  factor $\sqrt{N_c}$  results from taking transition matrix elements
between excited and ground state baryons \cite{Goity:2004pw}.
The reduced matrix elements
are expressed  in terms of the RME  of the spin-flavor
operator in ${\cal B}^{[LI]}$ by:
\bea
\langle  J^* I^*\parallel {\cal B}_a^{[LI]} \parallel N \rangle
\!\!\!&=&\!\!\!
\sqrt{(2 L + 1)(2 J^*+1) }
 \left\{
 \begin{array}{ccc}
 \fr 1 2 & \ell' & I^*\\
 \ell & J^* & L
 \end{array}
 \right\}
\nonumber \\
&  \times &\!\!\! (-1)^{L+\fr 12 +I^*+\ell} \langle S^*\!=\!I^*\!\!\parallel {\cal{G}}_a
\parallel \!\! N \rangle
\label{RME}
\eea
where the latter  RME is
evaluated using similar techniques to those in \cite{Goity,Carlson:1998vx}.
The coefficients $g_a(\omega)$ are determined  by fitting to the
empirical helicity amplitudes \cite{PDG06}. Their
$\omega$-dependencies are taken here to be the natural ones for
the multipole transitions, i.e., of the form
$g_a(\omega)=g_a\times(\omega/\Lambda)^n$, with $n=1$ for $M^{[1I]}$ and
$E^{[2I]}$, and $n=3$ for $M^{[3I]}$ and $E^{[4I]}$ operators.
In the fits below, given the typical values of $\omega$ involved,
a natural choice for $\Lambda$ turns out to be $\Lambda=m_\rho=770$ MeV.
Notice that the mass splittings within the baryon multiplets being
considered are ${\cal{O}}(1/N_c)$, and therefore one could  also absorb
the  $\omega$ dependencies  of the coefficients  into  coefficients of
higher order operators.

The sign $\eta(B^*)$  is obtained from the
strong amplitude  for $ B^*  \rightarrow  \pi N$ and is given in terms of
the sign of its  RME   as follows:
\beq
\eta(B^*) =(-1)^{J^*-\fr 12}\;
{\rm sign}(\langle \ell_\pi \; N\parallel H_{\rm QCD}\parallel J^*\;I^*\rangle),
\eeq
where $\ell_\pi$ corresponds to the pion partial wave.
The signs of the strong RME were determined in the $1/N_c$ expansion in
Ref. \cite{Goity:2005rg}.
That analysis can determine the signs up to an overall sign for
each pion partial wave. In the case of the Roper multiplet,
where only $\ell_\pi=1$ amplitudes contribute, this does not
bring in any ambiguity. However,  in the $[20,\, 2^+]$ multiplet  there is an undetermined overall relative
sign  between the RME of  $P$ and $F$ waves which has a bearing on relation (5). Following Ref.\cite{Babcock:1975bw}, we
introduce the notation $\xi'=\mbox{sign}(P/F)$ to refer to that relative  sign.

We turn now to  the fits of the empirical helicity
amplitudes.  Since the theoretical  errors  of a leading order (LO)  analysis are ${\cal{O}}(1/N_c)$,
in the LO fits we set the errors of the input amplitudes to be at least 30\%.
In  the next to leading order  (NLO)   fits we use the experimental errors,  which in some
cases are  around  10\%,    while most  are larger than that.

\begin{table}[h]
\begin{center}
\caption{Fit parameters $g_a$ and helicity amplitudes (in $10^3 \
{\rm MeV}^{-1/2}$) of  $[20,\ 0^+]$ baryons.
Errors are indicated in parenthesis.
}\label{tres}
\begin{tabular}{crrrrc}
\hline\hline
                  &  Emp.$\ \ $   &  LO   $\quad $   &  NLO1    &   NLO2 &$\eta$   \\
\hline
$\chi^2_{dof}$&             &  2.0   $\quad $  &    2.6  $\quad $  &     -   $\quad $ &\\
dof           &             &   3    $\quad $  &    2     $\quad $ &     0   $\quad $ &\\
\hline
$M_1^{[1,0]}$ &             &               &  $ 1.0 (0.4)$  & 0.8(0.4) & \\
$M_1^{[1,1]}$ &             & $2.5 (0.6)$ $\ $  &  $ 2.2 (0.3)$  & 2.3(0.3)&  \\
$M_2^{[1,1]}$ &             &               &                & 5.0(2.3)  &\\
$E_1^{[2,1]}$ &             &               &                &$-2.7(3.5)$  &\\
\hline
   $A^p_{1/2}[N(1440)]$
    & $-65(4)$  $\ $       &   $-33.4$   $\quad $    &   $-64.0$   $\quad $  &  $-64.7$ $\quad $& +1 \\
   $A^n_{1/2}[N(1440)]$
   &  $40(10)$  $\ $      &    $33.4$  $\quad $     &   $33.6$   $\quad $   &  $43.1$ $\quad $ & +1 \\
   $A^N_{1/2}[\Delta(1600)]$
   & $-23(20)$  $\ $      &   $-26.8$   $\quad $    &   $-31.3$  $\quad $   &  $-23.0$ $\quad $ & $-1$\\
   $A^N_{3/2}[\Delta(1600)]$
   & $-9(21)$   $\ $      &   $-46.3$   $\quad $    &   $-54.2$   $\quad $  &  $-9.0$  $\quad $ & $-1$\\
\hline \hline
\end{tabular}
\end{center}
\end{table}

 We first discuss the helicity amplitudes of the Roper
multiplet. Empirical amplitudes as well as the results of our fits
are displayed in Table \ref{tres}.
The two $\lambda=1/2$ amplitudes for the proton and
neutron $N(1440)$ are well known, while the two amplitudes
$\lambda=1/2$ and $3/2$  of the $\Delta(1600)$  have large errors.  At
leading order (LO) only the isotriplet operator $M^{[11]}_1$
contributes, and the fit gives $\chi^2_{\rm dof}=2$.
 The amplitudes for the $N(1440)$ require the
isosinglet contribution for a good fit, and also the LO result for
the $\lambda=3/2$ amplitude of  $\Delta(1600)$ is too large.
In the spirit of the SQTT,  the inclusion of  the
NLO isosinglet operator $M^{[10]}_1$ would be expected to
improve the fit. In fact, as seen from fit NLO1, this solves the first problem but
worsens the second one.
It is interesting to notice that in the LO fit the signs already coincide with the
empirical ones. The main  problem  is  the  small  magnitude  of the
$\lambda=3/2$ amplitude, especially because it is necessary a
  2-body operator  to fit it well.  As seen from fit NLO2, the operator
$M^{[11]}_2$ is the one that brings agreement while the  operator
$E^{[21]}_1$ turns out to be  almost insignificant as the relative error  of  its coefficient
shows. This seems to  be evidence that the electric quadrupole transition is
 small as   expected  if there is small mixing of  $\ell=2$ states in   the Roper multiplet.
A definite
conclusion requires, however,   an improvement in the empirical values
of the $\Delta(1600)$ amplitudes.  It is interesting to note that if one locks
the combination of 1-body operators
according to the  ratio of  isoscalar to isovector pieces of
the electric charge,   as one would do in a constituent quark model,
the fit remains consistent.  It is also interesting to note that a
different sign pattern for $\eta$  than the one resulting from the $1/N_c$
analysis of the strong transitions
\cite{Goity:2005rg}  leads  to a
considerably  worse fit at LO. This may be  indication of the
consistency between the $1/N_c$ analysis of strong vis-\`a-vis
electromagnetic transitions.

In  the   $[20,\, 2^+]$ multiplet,   the helicity amplitudes
associated with the $N(1680)$ and $\Delta(1950)$ are well known,
fairly known are those of the $\Delta(1920)$ and $\Delta(1905)$,
and finally the amplitudes of $N(1720)$ and $\Delta(1910)$ are
poorly established. This situation necessarily puts some
limitations on the significance of the analysis.  The empirical
amplitudes and  the results of our fits for the choice $\xi'=-1$
are displayed in Table \ref{cuatro}.
 At LO one has only 1-body operators, the isosinglet
$E^{[20]}_1$ and the three isotriplets: $M^{[11]}_1$,
$E^{[21]}_2$ and $M^{[31]}_1$.
\begin{table}[h]
\begin{center}
\caption{Fit parameters and helicity amplitudes (in $10^3 \
{\rm MeV}^{-1/2}$) of $[20,\,2^+]$ baryons.
Errors are indicated in parenthesis. Results for the choice $\xi'=-1$ are shown.}
\label{cuatro}
\begin{tabular}{crrrrr}
\hline\hline
                  &  PDG &  LO $\quad $    &  NLO1     &   NLO2  &$\eta$ \\
\hline
$\chi^2_{dof}$&             &  2.1 $\quad $   &          &     1.0  $\quad $ & \\
dof           &             &   11 $\quad $   &    0  $\quad $    &     9  $\quad $  &\\
\hline
$M_1^{[1,0]}$ &             &               &  $-0.9 (1.7)$  &             & \\
$M_1^{[1,1]}$&              & $-0.7 (0.5)$  &  $ 0.3 (1.0)$  &             & \\
$M_2^{[1,1]}$&              &               &  $ 2.3 (2.5)$  &             & \\
$M_3^{[1,1]}$&              &               &  $-1.9 (1.0)$  &             & \\
$E_1^{[2,0]}$&              & $0.3 (0.2)$   &  $ 1.2 (0.4)$  & $1.3(0.2)$  & \\
$E_2^{[2,0]}$&              &               &  $ 1.0 (1.3)$  &             & \\
$E_1^{[2,1]}$ &             &               &  $ 5.9 (1.9)$  & $7.1(1.4)$  & \\
$E_2^{[2,1]}$ &             & $0.0 (0.6)$   &  $ 1.6 (0.9)$  & $1.0(0.6)$  & \\
$E_3^{[2,1]}$ &             &               &  $ 8.8 (3.4)$  & $7.2(2.7)$  & \\
$E_4^{[2,1]}$ &             &               &  $ 0.9 (2.0)$  &             & \\
$M_1^{[3,0]}$ &             &               &  $ 3.6 (1.0)$  & $3.5(1.0)$  & \\
$M_1^{[3,1]}$ &             & $5.7 (0.9)$   &  $ 6.2 (0.6)$  & $6.0(0.4)$  & \\
$M_2^{[3,1]}$ &             &               &  $ 1.0 (2.4)$  &             & \\
$M_3^{[3,1]}$ &             &               &  $-0.3 (2.1)$  &             & \\
$E_1^{[4,1]}$ &             &               &  $-0.2 (2.5)$  &             & \\
\hline
   $A^p_{1/2}[N(1720)]$
    & $18(30)$        &   $1.0$  $\quad $     &   ~$18.0$  $\quad $   &  $47.2$  $\quad $& +1\\
   $A^p_{3/2}[N(1720)]$
    & $-19(20)$        &   $-15.1$  $\quad $     &   $-19.0$   $\quad $  &  $-27.2$ $\quad $ & +1\\
   $A^n_{1/2}[N(1720)]$
   &  $1(15)$        &    $13.0$  $\quad $     &   ~ $1.0$   $\quad $   &  $11.7$ $\quad $ & +1 \\
   $A^n_{1/2}[N(1720)]$
   &  $-29(61)$        &    $7.0$  $\quad $     &   $-29.0$   $\quad $   &  $-6.8$ $\quad $ & +1 \\
   $A^p_{1/2}[N(1680)]$
    & $-15(6)$        &   $-24.1$  $\quad $     &   $-15.0$   $\quad $  &  $-15.1$ $\quad $ & +1 \\
   $A^p_{3/2}[N(1680)]$
    & $133(12)$        &   $28.7$  $\quad $     &  ~ $133.0$  $\quad $   &  $127.8$ $\quad $ & +1 \\
   $A^n_{1/2}[N(1680)]$
   &  $29(10)$        &    $35.4$  $\quad $     &   ~ $29.0$  $\quad $    &  $25.7$  $\quad $& +1  \\
   $A^n_{1/2}[N(1680)]$
   &  $-33(9)$        &    $-12.7$  $\quad $     &   $-33.0$   $\quad $   &  $-34.3$  $\quad $& +1 \\
   $A^N_{1/2}[\Delta(1910)]$
   & $3(14)$        &   $-13.8$  $\quad $     &  ~ $3.0$   $\quad $  &  $0.0$ $\quad $ & +1 \\
   $A^N_{1/2}[\Delta(1920)]$
   & $40(14)$        &   $7.3$  $\quad $     &  ~ $40.0$  $\quad $   &  $16.3$  $\quad $& $-1$ \\
   $A^N_{3/2}[\Delta(1920)]$
   & $23(17)$         &   $12.0$  $\quad $     & ~  $23.0$  $\quad $   &  $-9.4$  $\quad $& $-1$  \\
   $A^N_{1/2}[\Delta(1905)]$
   & $26(11)$        &   $26.4$  $\quad $     &   ~ $26.0$  $\quad $   &  $21.1$  $\quad $& +1 \\
   $A^N_{3/2}[\Delta(1905)]$
   & $-45(20)$         &   $-19.3$  $\quad $     &   $-45.0$  $\quad $   &  $-49.5$  $\quad $& +1  \\
   $A^N_{1/2}[\Delta(1950)]$
   & $-76(12)$        &   $-55.2$  $\quad $     &   $-76.0$  $\quad $   &  $-77.3$  $\quad $ & $-1$\\
   $A^N_{3/2}[\Delta(1950)]$
   & $-97(10)$         &   $-71.2$  $\quad $     &   $-97.0$  $\quad $   &  $-99.8$  $\quad $ & $-1$ \\
\hline \hline
\end{tabular}
\end{center}
\end{table}

 This  fit leads to a reasonably good description
of the $\Delta(1950)$ amplitudes, which  are entirely given by the operator $M^{[31]}_1$.

This holds true at NLO where
they receive only small  contributions from other $M_3$ and the $E_4$ operators.
The main   contribution  to the $\chi^2$ is due to the large $\lambda=3/2$ amplitude
of the proton $N(1680)$, which is badly underestimated. It receives contributions from
the $M^{[31]}_1$ and the $E_2$ operators $E_1^{[20]}$ and $E_2^{[21]}$, but they are not
sufficient to give a good description.  This
is a well known problem which shows up in virtually every model
that has been considered, in particular constituent quark models.
 As the errors of their coefficients
show, at LO the  dominant operator is  $M^{[31]}_1$, while the
operator  $E^{[21]}_2$  contributes  among others to the
 large    $\lambda=3/2$ amplitude of  $N(1680)$.  The $M^{[11]}_1$ operator has a small coefficient
 which is a manifestation of  the general fact that   $M_1$ amplitudes are small in the $[20,\ 2^+]$ multiplet.  It should be noted that the choice $\xi'=+1$ leads to a
qualitative similar LO fit.
The NLO analysis is somewhat limited
by the  large errors of the inputs,  which exceed in general the
10\% error  that would allow for an accurate NLO analysis.  The
NLO1 fit shows  that  for $\xi'=-1$  the coefficients  $g_a$ needed
to reproduce the empirical amplitudes are all of natural
magnitude,   indicating a good convergence of the $1/N_c$
expansion.   It also
implies that a reduced number of operators give the significant
contributions as shown by the relative errors of the coefficients.
This is confirmed by the NLO2 fit where we have included the
minimum number of operators that allow for a $\chi^2_{\rm dof}
\simeq 1$. It is very important  to note that a fit that keeps
only 1-body operators among the significant operators leads to a
$\chi^2_{\rm dof}\sim 2$, which is showing  that the dominant
effects result from the coupling of the photon to a single quark.
The case $\xi'=+1$  gives  a larger number of
relevant operators with the coefficients corresponding to some of the electric quadrupole operators
turning out to be unnaturally large. We thus conclude
  that the choice $\xi'=-1$ is favored by the
present $1/N_c$ analysis. The old SQTT
analysis\cite{Babcock:1975bw} also indicated a preference for
$\xi'=-1$.
Returning to  the large $\lambda=3/2$ amplitude of the
$N(1680)$,  at NLO it receives contributions from  only 1-body
operators, namely from both isosinglet $E_2$ operators,  from
$E_1^{[21]}$ and $E_2^{[21]}$, and from the $M_3$ operators
$M_1^{[31]}$ and  $M_1^{[30]}$. All these contributions have  the
same sign and none is dominating, which  makes  the  understanding
of the large magnitude of the amplitude difficult.    
  Most amplitudes receive several contributions, and there are large
cancellations taking place, in particular in small amplitudes. One
exception is the magnetic dipole amplitude  $\lambda=1/2$ of the
$\Delta(1910)$ which is experimentally very small, and in this
analysis it receives a single 1-body operator contribution, namely
that of the operator $M^{[11]}_1$. This serves to set  the
benchmark for the magnitude of magnetic dipole amplitudes, which
as the fits show are  small. Finally,  the only $E_4$ operator
present in the analysis turns out to be irrelevant, giving
insignificant contributions to the  $\Delta(1950)$ amplitudes.

 The results obtained here show the dominance of $M_1$
transitions in the $[20,\, 0^+]$ and of the $M_3$ and $E_2$
transitions in the $[20,\, 2^+]$.  In addition,  although the fits
cannot establish with certainty the relevance of 2-body effects,
they give  some evidence of their presence.
This should motivate the  consideration
of  mechanisms that could give rise to these 2-body  effects, some of which have been proposed
 by several authors \cite{Exchange} in the context of quark models.

We would like to thank I. Aznauryan, S. Capstick, T. S. H. Lee and
W. Roberts for useful discussions.
This work was supported by DOE (USA) through contract DE-AC05-84ER40150,
by the NSF (USA) through grants \#~PHY-9733343 and -0300185, and by CONICET (Argentina)
grant \# PIP 6084 and by ANPCyT  (Argentina) grant \# PICT04 03-25374.

\end{document}